\documentclass[twocolumn,superscriptaddress]{revtex4}
\usepackage{amssymb}
\usepackage{amsmath} 
\usepackage{amsbsy}
\usepackage[dvips]{graphicx}
\usepackage{multirow}
\usepackage{natbib}

\newcommand{\Tr}{\mathrm{Tr}}

\newcommand{\Id}{\ensuremath{\mbox{I\hspace{-.2em}I}}}

\newlength{\barheight}
\newcommand{\Prob}[1]{\ensuremath{\mathrm{Prob}\left(#1\right)}}

\begin{document}

\title{Progress toward scalable tomography of quantum maps using twirling-based methods and information hierarchies}
\author{Cecilia C. L\'{o}pez}
\affiliation{Department of Nuclear Science and Engineering, MIT, Cambridge, Massachusetts 02139, USA}
\affiliation{Theoretische Physik, Universit\"{a}t des Saarlandes, D-66041 Saarbr\"{u}cken, Germany}
\affiliation{Departament de F\'{i}sica, Universitat Aut\`{o}noma de Barcelona, E-08193 Bellaterra, Spain}
\author{Ariel Bendersky}
\affiliation{Departamento de F\'{i}sica, FCEyN, UBA, Ciudad Universitaria, 1428 Buenos Aires, Argentina}
\author{Juan Pablo Paz}
\affiliation{Departamento de F\'{i}sica, FCEyN, UBA, Ciudad Universitaria, 1428 Buenos Aires, Argentina}
\author{David G. Cory}
\affiliation{Department of Nuclear Science and Engineering, MIT, Cambridge, Massachusetts 02139, USA}
\affiliation{Perimeter Institute for Theoretical Physics, Waterloo, Ontario N2J 2W9, Canada}

\begin{abstract}
We present in a unified manner the existing methods for scalable partial
quantum process tomography. We focus on two main approaches: the one
presented in Bendersky et al. [Phys. Rev. Lett. 100, 190403 (2008)], and the ones described,
respectively, in Emerson et al. [Science 317, 1893 (2007)] and L\'{o}pez et al. [Phys. Rev. A 79, 042328 (2009)], 
which can be combined together. The methods share an essential
feature: They are based on the idea that the tomography of a quantum map
can be efficiently performed by studying certain properties of a
twirling of such a map. From this perspective, in this paper we present extensions,
improvements and comparative analyses of the scalable methods for
partial quantum process tomography. We also clarify the significance of the
extracted information, and we introduce interesting and
useful properties of the $\chi$-matrix representation of quantum maps
that can be used to establish a clearer path toward achieving full
tomography of quantum processes in a scalable way.
\end{abstract}

\maketitle
\section{Introduction}
\label{sec:intro}

The number of parameters describing a quantum map scale exponentially with $\ln(D)$, 
with $D$ the dimension of the Hilbert space $\mathcal{H}_D$ of the system.
One can then argue that the resources required to obtain this exponentially large number of parameters
will also necessarily increase exponentially. This is why the complete characterization of
a quantum map is considered to be a \textit{nonscalable} task. The task of characterizing a quantum map
is known as quantum process tomography (QPT) \cite{qpt} and the above is the main reason why full QPT
is exponentially expensive. Moreover, many existing methods have another major defect as they
are inefficient also in extracting partial information about the quantum process
(for a review, see \cite{mohseni_rezakhani}).
Recently, however, several works
\cite{emerson_alicki, levi_lopez,dankert,emerson_silva,silva_magesan,bendersky_paz,lopez_levi,bendersky_paz_2}
have demonstrated that it is possible to extract partial but nevertheless relevant information
in an efficient way [where by efficient we mean that it is done at a cost that scales at most
polynomially with $\ln(D)$)]. This has opened a new chapter in quantum information processing
toward the scalable characterization of quantum processes. These new methods share a common feature.
They are based on the idea that the relevant properties of the quantum map can be obtained by averaging properties
of a family of maps which are obtained from the original one. The averaging is done by an operation denoted as
twirling \cite{twirling} (which will be defined in detail later) and involves the application of certain operations
before and after the application of the map.\\

In this work we present a review of the recent methods for partial QPT, establishing
 connections between them and adding results. We not only present a unifying perspective of these methods but also
develop a better understanding of the problem at hand -- the tomographic
characterization itself. 

The paper is organized as follows: In Sec. \ref{sec:chi} we introduce
the $\chi$-matrix description of a quantum process distinguishing completely positive (CP) maps and others
that are not CP. In Sec. \ref{sec:twirl} we present the basic ideas behind the notion of a twirling operation.
We show that the elements of the $\chi$-matrix can be obtained from this type of operation.
Moreover, some important properties of such a matrix (in particular, some useful relations between diagonal
and off-diagonal elements) are discussed in Secs. \ref{subsec:CP} and \ref{subsec:P}.

In Sec. \ref{sec:fullTwirl} we review the method of ``selective efficient quantum process tomography,''
originally presented in \cite{bendersky_paz,bendersky_paz_2}.
We reformulate this approach by using more general types of twirlings. Not only do we highlight
the power of this method but also we establish the convenience of one type of twirl over another.
Furthermore, we provide a clear prescription for its implementation when targeting the scalable
measurement of several $\chi$-matrix elements at a time.

In Sec. \ref{sec:1qubitTwirl} we move to protocols utilizing simpler forms of twirling, which are
substantially less demanding regarding their experimental implementation.
We take the results from \cite{emerson_silva} and \cite{lopez_levi} and present them in a new compact form
as a single protocol enabling us to obtain the diagonal elements of the $\chi$-matrix
grouped by ``how many'' and ``which'' qubits are affected by the quantum map. By fully proving the
method by construction, we aim to further clarify its simple implementation as well as its limitations.

Finally, in Sec. \ref{sec:diagonal}, we discuss the potential of these strategies toward achieving scalable complete
tomography of a quantum process. We believe that the key to this lies in the hierarchization
of the exponentially large number of parameters (in which the results of Secs. \ref{subsec:CP} and \ref{subsec:P} play an important role).
The methods described in Secs. \ref{sec:fullTwirl} and \ref{sec:1qubitTwirl} retrieve the diagonal elements
of the $\chi$-matrix, and in Sec. \ref{sec:diagonal} we show how the diagonal elements provide information not only about themselves
but also about the off-diagonal ones. This is what we identify as an information hierarchy.

Furthermore, we hope that this article sets a practical path for experimentalists looking to implement
quantum process characterization in a quantum information setting, that is, when the scalability of the tomographic method matters.

\section{The $\chi$-matrix description of a quantum process}
\label{sec:chi}

A general quantum process can be described by the action of an arbitrary map $\Lambda$ on the state $\rho$ in $\mathcal{H}_D$.
Any linear map $\Lambda$ can be expressed as
\begin{eqnarray}
\mathbf{\Lambda}(\rho) &=& \sum_{l,l'=0}^{D^2-1} \chi_{l,l'} E_l \rho E_{l'}^\dag
\label{chimatrixE}
\end{eqnarray}
where the operators $\{E_l, l=0, \dots, D^2-1\}$ form a basis for the space of operators $\mathcal{H}_D$.
The complex numbers $\chi_{l,l’}$ form the so-called
$\chi$-matrix of the map. The $\chi$-matrix is obviously dependent on the operator basis.
Without loss of generality, we can take this basis to be orthogonal, that is, to be such that $\Tr{[E_l^\dag E_{l'}]}=D \delta_{l,l'}$ \cite{basisE}.
It is simple to show that the map preserves the hermiticity of $\rho$ if and only if the $\chi$ matrix is Hermitian itself
(i.e., if $\chi_{l,l'}=\chi_{l',l}^*$).
Moreover, the map $\Lambda$ is trace-preserving if and only if the condition
$\sum_{l,l'} \chi_{l,l'} E_{l'}^\dag E_{l} =I$ is satisfied. In such a case it is simple to count the number of independent real parameters
defining the quantum map, which turns out to be $D^4-D^2$ (with the trace preserving condition implying a reduction of the number of
parameters in $D^2$ and also implies that the condition $\sum_{l} \chi_{l,l} =1$ must be satisfied).

We remark that this description is valid for any linear map. For the case of Hermitian maps it is possible to uncover further structure.
In such a case the $\chi$-matrix can be diagonalized by a unitary transformation.
In matrix notation we can write $\chi=B^\dagger S B$, where $S$ is a diagonal matrix with real eigenvalues.
The columns of the unitary matrix $B$ define the eigenvectors: The $m$-th component of the $l$-th eigenvector $\bar{b}_l$
is $(\bar{b}_l)_m=B_{m,l}$.
By using this notation it is evident that the elements of the $\chi$ matrix can be obtained as
$\chi_{l,l'}=\bar{b}_l^\dagger S \bar{b}_{l'}=\sum_m B_{m,l}^*S_{m,m}B_{m,l'}$. \\

Some simple but useful results follow from this expression. Replacing it in the original formula for the map given
in Eq. (\ref{chimatrixE}) we obtain the following alternative expression for an arbitrary linear Hermitian map:
\begin{eqnarray}
\mathbf{\Lambda}(\rho) &=& \sum_{k} S_{k,k} A_k \rho A_{k}^\dag
\label{kraus}
\end{eqnarray}
where the operators $A_k$ form an orthonormal basis defined as $A_k=\sum_l B_{k,l}^* E_l$. (The orthonormality of $A_k$ follows
from the fact that these operators are a linear combination of the original $E_l$ with coefficients that are elements of a unitary matrix.)
It is worth noticing that the coefficients $S_{kk}$, which are the eigenvalues of the $\chi$-matrix, are necessarily real but can be
either positive or negative. This representation for the quantum map is closely related to the so-called Kraus representation
(which is obtained only if the eigenvalues $S_{mm}$ are all positive, which is in turn valid for the case of CP maps only, as discussed below).
In fact, Eq. (\ref{kraus}) is a generalization of the Kraus representation valid for any linear Hermitian map.

More generally, the above expressions make evident that an arbitrary linear Hermitian map can always be written as the difference
between two CP maps \cite{shabani_lidar} (this is the case since any matrix $S$ can be expressed as the difference between two positive matrices).\\

\subsection{$\chi$-matrix of completely positive maps}
\label{subsec:CP}

Using the above results, we can derive some properties for the $\chi$-matrix of completely positive maps
(i.e., when the map $\Lambda$ and any trivial extension of it to a bigger Hilbert space preserve positivity).
Since the matrix $S$ is positive, it is clear that matrix elements $\chi$ are obtained as the inner product between
two eigenvectors $\bar{b}_l$ defined through the positive matrix $S$,
\begin{align}
\langle \bar{b}_l, \bar{b}_{l'}\rangle \equiv \bar{b}_l^\dagger S \bar{b}_{l'}
\notag
\end{align}
From this observation we can conclude the following: First, it is evident that diagonal elements must be positive
(i.e., $\chi_{l,l} \geq 0 \ \forall \ l$).
Moreover, as the inner product satisfies the Cauchy-Schwarz inequality,
we can obtain the following relation between diagonal and off-diagonal elements of the $\chi$-matrix:
\begin{eqnarray}
|\chi_{l,l'}|^2 \leq \chi_{l,l} \ \chi_{l',l'}
\label{boundCP}
\end{eqnarray}
\textit{This means that, for any CP map, the diagonal elements of the $\chi$-matrix are always nonnegative, and that they bound the corresponding off-diagonal elements.}
These two simple results are quite significant and they will prove very useful later on.
Below, we will derive another relation between diagonal and off-diagonal coefficients for the $\chi$-matrix valid for positive (not necessarily CP) maps.
In this way we will be also able to establish some conditions to distinguish these two important classes of maps.

\section{Twirling of a map, and sampling of a twirl}
\label{sec:twirl}

\begin{figure}[h]
\includegraphics[width=0.37\textwidth, angle=0]{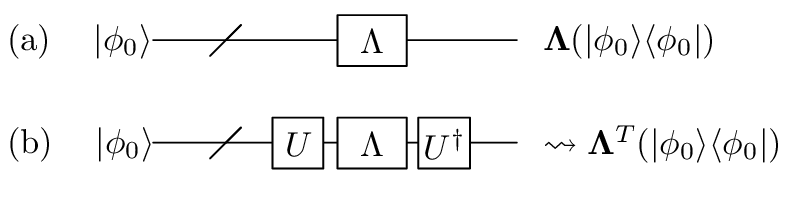}
\caption{Circuit representation of (a) the action of a map $\Lambda$; (b) the action of the map, now twirled by $U$.}
\label{fig:twirling}
\end{figure}

The action of twirling a map is depicted in Fig. \ref{fig:twirling}.
We have a quantum process characterized by a map $\Lambda$ that acts on a system (for example, a quantum information processor)
originally prepared in an arbitrary state $|\phi_0\rangle$, as depicted in Fig. \ref{fig:twirling}(a). We twirl the map by applying an operator
$U$ before the map, and an operator $U^\dag$ after, as in Fig. \ref{fig:twirling}(b). Typically the twirling is considered
as the average of this over different elements $U$, resulting in a net map $\Lambda^\mathrm{T}$, the twirled map.
Different families of $U$'s will return different types of twirl.

In particular, we are initially interested in the Haar twirl,
\begin{eqnarray}
\mathbf{\Lambda}^{HT}(\rho) &=& \int dU U^\dag \mathbf{\Lambda}(U \rho U^\dag) U
\label{haarU}
\end{eqnarray}
where $dU$ denotes the unitarily invariant Haar measure on U($D$).

There is a version of this twirl where the average is over the Haar measure in state space,
\begin{eqnarray}
\langle \phi_0 | \mathbf{\Lambda}^{HT}(|\phi_0\rangle \langle \phi_0|)|\phi_0\rangle = \int d\psi \langle \psi| \mathbf{\Lambda}(|\psi\rangle \langle \psi|)|\psi\rangle
\label{haarState}
\end{eqnarray}
The relation between the two is straightforward if we notice that if $U$ is randomly
drawn according to the Haar measure on operator space, then $|\psi\rangle = U|\phi_0\rangle$
corresponds to the Haar measure on vector space -- for any arbitrary fixed state $|\phi_0\rangle$.

There are several previous results concerning the Haar twirl, in its forms both in operator space \cite{haartools,emerson_alicki}
and in state space \cite{caves}. Summarizing this literature, we limit ourselves to state the following general mathematical formula:
\begin{align}
& \int dU \Tr[A_1 U^\dag B_1 U A_2 U^\dag B_2 U] \notag \\
&= \frac{\Tr[A_1 A_2]}{D^2-1} \left(\Tr[B_1]\Tr[B_2] -\frac{\Tr[B_1 B_2]}{D}\right) \notag \\
&+ \frac{\Tr[A_1]\Tr[A_2]}{D^2-1} \left(\Tr[B_1 B_2] -\frac{\Tr[B_1]\Tr[B_2]}{D}\right)
\label{generalHaarTwirl}
\end{align}
for any operators $A_1$, $A_2$, $B_1$, $B_2$ in $\mathcal{H}_D$.

Given this and explicitely using the trace-preserving condition, the $\chi$-matrix elements can be expressed as the outcome of a twirl,
\begin{eqnarray}
\frac{D \chi_{l,l'} + \delta_{l,l'}}{D+1} = \int d\psi \langle \psi| \mathbf{\Lambda}(E_l^\dag |\psi\rangle \langle \psi| E_{l'})|\psi\rangle
\label{main_ariel}
\end{eqnarray}
as already stated in \cite{bendersky_paz}.

\subsection{The $\chi$-matrix of positive (but not necessarily CP) maps}
\label{subsec:P}

Equation (\ref{main_ariel}) is valid for any map $\Lambda$ under study. In particular, for processes that
take positive operators into positive operators, Eq. (\ref{main_ariel}) defines a valid inner product
\begin{align}
\langle E_l,E_{l'}\rangle \equiv \int d\psi \langle \psi| \mathbf{\Lambda}(E_l^\dag |\psi\rangle \langle \psi| E_{l'})|\psi\rangle
\notag
\end{align}
Notice that we have $\langle E_l,E_l\rangle = (D \chi_{l,l} + 1)/(D+1) \geq 0$.
This implies that for a positive (but not necessarily CP) map, we have that the diagonal elements
of the $\chi$-matrix can be negative but only up to an exponentially vanishing value: $\chi_{l,l} \geq -1/D$.
Also, notice that $\langle E_l,E_l\rangle$ is a survival probability: the probability
of the system remaining in its initial state after applying the twirled map $\Lambda^{\mathrm{HT}}$ to it.
Therefore, $(D \chi_{l,l} + 1)/(D+1) \leq 1$, which implies $\chi_{l,l} \leq 1$.\\

Moreover, again using the Cauchy-Schwarz inequality on this inner product, we obtain that
for $l \neq l'$
\begin{eqnarray}
|\chi_{l,l'}|^2 \leq \chi_{l,l} \ \chi_{l',l'} + \frac{\chi_{l,l} + \chi_{l',l'}}{D} + \frac{1}{D^2}
\label{boundP}
\end{eqnarray}
So for large systems where we can consider $D >> 1$, the off-diagonal matrix elements are effectively bound by the diagonal ones.
These bounds also suggest that non-CP but positive processes are ``exponentially close'' to CP ones. This is an interesting
result in the framework of open quantum systems, where there are still important discussions about what mathematical
conditions a physical map should fulfill \cite{shabani_lidar,pechukas_alicki,carteret}.

\subsection{(Approximate) sampling of a twirl}

\begin{figure}[h]
\includegraphics[width=0.37\textwidth, angle=0]{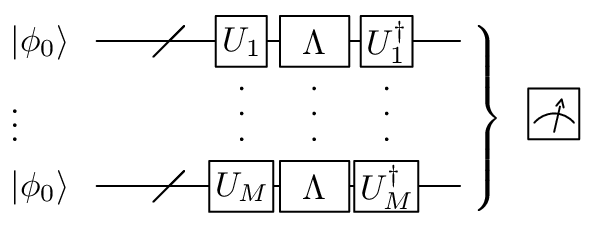}
\caption{Circuit representation of the twirling of $\Lambda$, approximated by sampling $M$ times over the elements that constitute the family of
twirl operators (the $U$'s). Each time the system is prepared in the same initial state $|\phi_0\rangle$.
The average of these $M$ measurements will retrieve the desired probabilities.}
\label{fig:twirling2}
\end{figure}

Equation (\ref{main_ariel}) already demonstrates the usefulness of twirls in extracting the elements of the $\chi$-matrix.
This is indeed what lies at the heart of the methods developed in \cite{emerson_alicki,levi_lopez,dankert,emerson_silva,bendersky_paz,lopez_levi,bendersky_paz_2}.
It is evident then that we will need to implement the twirl experimentally (in either operator or state space).
Unfortunately, as we will see, the number of
elements in the twirls that are of our interest is infinite or grows exponentially with $\ln(D)$ \cite{gross_eisert}. Thus implementing
the twirl perfectly is a nonscalable task.
However, it was initially suggested in \cite{emerson_alicki,levi_lopez} that we can approximate the twirl by sampling randomly over the
family of $U$'s, say $M$ times, as depicted in Fig. \ref{fig:twirling2}.
In the case of a state twirl, the approach would be the same, since in practice a state twirl results
from implementing a series of operations (which would take the place of the $U$'s) on a convenient initial state $|\phi_0\rangle$.

If we are interested in
measuring the probability of finding the system in any given state,
this outcome will be a boolean variable retrieved with a standard deviation
$\sigma \leq 1/\sqrt{M}$ (following the central limit theorem with $M \rightarrow \infty$), so for a desired precision $\epsilon$ we must have $M \geq \epsilon^{-2}$.
On the other hand, the Chernoff bound tells us that
for a desired precision $\epsilon$ and an error probability $\delta \ll 1$,
we must have $M \geq \ln(2/\delta)/(2\epsilon^2)$, which is a stronger requirement when $\delta < 2 e^{-2}$.
This is a bound to the error probability and not to the error itself, however it is rigorous for arbitrary $M$.
In any case $M$ should satisfy both conditions \cite{lopez_thesis}. Since $M$ is independent of the size of the system, this ensures the scalability of the
 experimental implementation if each of the $M$ realizations themselves can be implemented efficiently.
This holds of course unless the targeted probabilities are expected to be of the $O(1/\sqrt{D})$,
in which case the estimation of each probability would require an exponentially large number of realizations. However this would be the case of a process
close to a random channel, and usually
they are of no interest in quantum information
and/or in relatively controllable quantum systems.

Finally, we note that we could separate the average of binary outcomes (the result of projective measurements) required to determine the probability
for an experiment with a fixed twirl operator $U$, from the average of experiments with different $U$'s. This is useful in cases where
repeating an experiment with a fixed $U$ is trivial compared to running a new one with a different twirl operator.
In this case, other interesting bounds
to the error can be applied, as for example in the experimental work in \cite{schmiegelow_larotonda}.\\

In what follows we restrict ourselves to a Hilbert space that is an $n$-fold tensor product
of a two-level system space, so $D=2^n$. Moreover, we work with a specific set of operators $\{ E_l \}$: the generalized
Pauli operators (also called the product operator basis). We will specifically denote them as $\{ P_l \}$,
$P_l = \bigotimes_{j=1}^n P_l^{(j)}$.
Each $P_l^{(j)}$ is an element of the Pauli group $\{I, \sigma_x, \sigma_y, \sigma_z\}$ for the $j$-th qubit.
$P_0=\Id$ is the identity operator
in $\mathcal{H}_D$, and for $l >0$ at least one factor in each $P_l$ is a Pauli matrix. Notice that $P_l^\dag = P_l$ and that $\Tr[P_l P_{l'}]=D \delta_{l,l'}$ indeed.
From now on, the $\chi$-matrix elements will be always associated to this basis.

\section{Methods using a full space twirling of the map under study}
\label{sec:fullTwirl}

In this section we start by studying the methods utilizing a full twirl over U($D$).
If the twirl depicted
in Fig. \ref{fig:twirling2} is over U($D$), the survival probability is the average fidelity of the original
map $\Lambda$ \cite{dankert,bendersky_paz,bendersky_paz_2}.
This is in fact Eq. (\ref{haarState}), which is the definition of average fidelity
of a quantum channel $\Lambda$ \cite{nielsen}, $F(\Lambda)$.

The tomographic methods in \cite{bendersky_paz,bendersky_paz_2} are actually presented
not in terms of twirl operators
but rather in terms of the states of mutually unbiased bases (MUBs): $\{ |\psi_{J,m}\rangle, J=0,\ldots D;\ m=1,\ldots D \}$.
Here we introduce their equivalents
using twirls in operator space. Nevertheless, further analysis will in turn lead to a slight preference toward the former one.

We rely on \cite{dankert} to establish the equivalence between the Haar twirl in operator space and
the Clifford twirl in U($D$) for $D > 2$ (and we will explicitly prove it for $D=2$ in Sec. \ref{sec:1qubitTwirl}).
On the other hand, the equivalence between a Haar twirl in state space and a twirl using MUB states,
for dimensions that are powers of prime numbers, is presented in \cite{paper_MUB}. Altogether, we can write
\begin{align}
&\langle \phi_0| \mathbf{\Lambda}^{\mathrm{HT}}(|\phi_0\rangle \langle \phi_0|) |\phi_0\rangle = \nonumber\\
&= \frac{1}{|\mathcal{C}|} \sum_{l=1}^{|\mathcal{C}|} \langle \phi_0| \mathcal{C}_l^\dag \mathbf{\Lambda}
(\mathcal{C}_l |\phi_0\rangle \langle \phi_0| \mathcal{C}_l^\dag) \mathcal{C}_l |\phi_0\rangle \nonumber\\
&=\frac{1}{D(D+1)} \sum_{J,m} \langle \psi_{J,m}| \mathbf{\Lambda}(|\psi_{J,m}\rangle \langle \psi_{J,m}|) |\psi_{J,m}\rangle
\label{MUB_CDT}
\end{align}
where the $\mathcal{C}_l$ are the Clifford operators in U($D$) and $|\phi_0\rangle$ is an arbitrary fixed state.
Both these twirls imply the same cost, as preparing MUB states starting from the computational basis and implementing
the $\mathcal{C}_l$ require the same resources: $O(n^2)$ one-qubit and two-qubit gates \cite{bendersky_paz_2,dankert_thesis,stabilizers}.
And again, the number of Clifford operators $|\mathcal{C}|$ scales exponentially with $\ln(D)$, as does the number
of MUB states (so in both cases we will resort to sampling the twirl).

\begin{figure}[h]
\centering
\includegraphics[width=0.37\textwidth, angle=0]{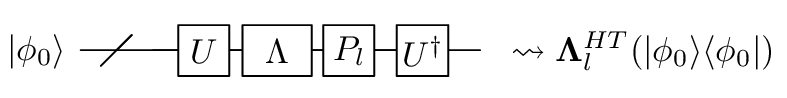}
\caption[Circuit representation a twirled $\Lambda_l$ with
$\Lambda_l(\rho)=P_l \mathbf{\Lambda}(\rho) P_l$.]{Circuit representation of the action of a map $\Lambda_l$ with
$\Lambda_l(\rho)=P_l \mathbf{\Lambda}(\rho) P_l$, twirled by $U$.}
\label{fig:twirling3}
\end{figure}

In \cite{bendersky_paz,bendersky_paz_2} it was shown how to selectively measure any diagonal $\chi$-matrix element using an MUB twirl.
There is an equivalent
to this using a Clifford twirl in U($D$). As presented in \cite{bendersky_paz},
if we implement an intermediary extra gate $P_l$ before completing the twirl (see Fig. \ref{fig:twirling3}),
the survival probability is
\begin{align}
\Tr[|\phi_0\rangle \langle \phi_0| \mathbf{\Lambda}_l^{HT}(|\phi_0\rangle \langle \phi_0|)] = \frac{D \chi_{l,l} + 1}{D+1}
\label{survival_m}
\end{align}
This can be proven straightforwardly from Eq. (\ref{main_ariel}).

We are thus able to measure efficiently one $\chi_{l,l}$ at a time ($selective$ efficient quantum process tomography - SEQPT \cite{bendersky_paz}).
However, we can modify the protocol to automatically select and retrieve the largest $\chi_{l,l}$: the coefficients
such that $\chi_{l,l} \geq 2/M$. The strategy goes as follows.\\

\begin{figure}[h]
\includegraphics[width=0.37\textwidth, angle=0]{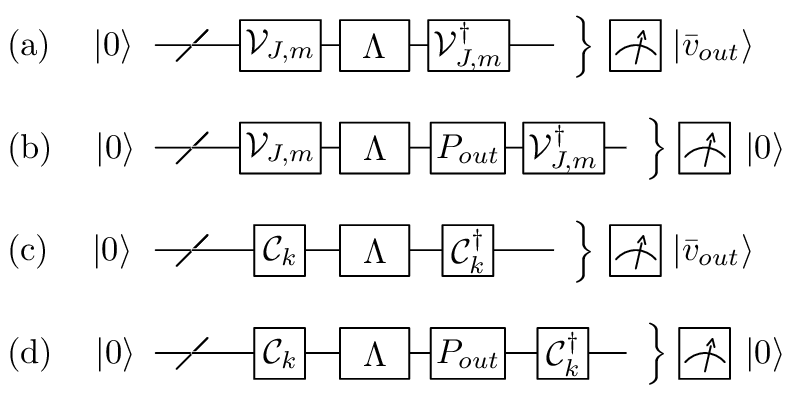}
\caption{Circuit representation of equivalent schemes to determine the largest $\chi_{l,l}$, by considering pairs
of experiments.}
\label{fig:inPairs}
\end{figure}

We first revisit the method as presented in \cite{bendersky_paz} for an MUB twirl.
As depicted in Fig. \ref{fig:inPairs}(a), we consider a single experiment where the system is
prepared in a randomly chosen MUB state $|J,m\rangle = \mathcal{V}_{J,m}|0\rangle$. $\mathcal{V}_{J,m}$
represents the change of basis operation between the computational state $|0\rangle$ and the targeted MUB state.
We measure at the end the state in the computational (Zeeman) basis, obtaining then an $n$-bit string $|\bar{v}_{out}\rangle$
(where $\bar{v}_{out}$ is a boolean vector of length $n$ that labels the states as binary numbers).
Considering that $\mathcal{V}_{J,m}|\bar{v}_{out}\rangle = |J,m'\rangle$ is just another state of the MUB, and that there are
$D$ possible Pauli operators that take $|J,m\rangle$ to $|J,m'\rangle$ (up to a global phase) \cite{bendersky_paz_2},
we can regard this experiment as equivalent to the one in Fig. \ref{fig:twirling3}, but where now we have $D$ possible Pauli operators playing
the role of the intermediary $P_l$.

To gain further insight into the mechanism of this result, we recall these dynamics using the stabilizer formalism \cite{stabilizers}.
We describe the state $|\bar{v}_{out}\rangle$ with the
subset $\mathcal{B}_Z$ formed by the $n$ commuting Pauli operators $\{ \sigma_z^{(1)}, \ldots, \sigma_z^{(n)}\}$ and a string $\bar{s}_{out}$ of $n$ signs, $\pm 1$,
corresponding to the eigenvalues
of $|\bar{v}_{out}\rangle$ for that subset.
These $n$ operators generate the maximally commuting (Abelian)
group of $D$ Pauli operators that stabilize the computational basis.
On the other hand, the state $|0\rangle$ is described by $\mathcal{B}_Z$ with a string $\bar{s}_{0}$ of all $+1$ signs.
The action of $\mathcal{V}_{J,m}$ on $|0\rangle$ is equivalent to changing $(\mathcal{B}_Z, \bar{s}_{0})$ to $(\mathcal{B}_J, \bar{s}_{0})$, where now $\mathcal{B}_J$ is another
subset of $n$ commuting Pauli operators -- the generators of the group that stabilizes the $D$ states corresponding to the MUB labeled by $J$.
Also, the action of $\mathcal{V}_{J,m}$ on $|\bar{v}_{out}\rangle$ is equivalent to changing $(\mathcal{B}_Z, \bar{s}_{out})$ to $(\mathcal{B}_J, \bar{s}_{out})$.
We now use that the state $(\mathcal{B}_J, \bar{s}_{out})$ can be thought as the result of a Pauli operator $P_{out}$ acting on $(\mathcal{B}_J, \bar{s}_{0})$, which leaves us with
the scheme depicted in Fig. \ref{fig:inPairs}(b). $P_{out}$ must fulfill the requisite of commuting (anticommuting) with the Pauli operators in $\mathcal{B}_J$ 
that have a corresponding $+1$ ($-1$) in $\bar{s}_{out}$. We express this condition as
the commutation relations
\begin{align}
[P_{out},\mathcal{V}^\dag_{J,m} \sigma_z^{(j)} \mathcal{V}_{J,m} ]_{\pm} = 0, \ \ \ j=1, \ldots, n
\label{thecommutator}
\end{align}
where the $[\ ,\ ]_\pm$ stands for commutator or anticommutator, depending on the signs of $\bar{s}_{out}$.
But as already stated before, there will be $D$ possible candidates for the intermediary $P_{out}$.
This can be seen as follows. First, we notice that Eq. (\ref{thecommutator}) can be rewritten as
$[ P'_{out}, \sigma_z^{(j)} ]_{\pm}= 0$
where we have defined
\begin{align}
P'_{out} \equiv \mathcal{V}_{J,m} P_{out} \mathcal{V}^\dagger_{J,m}
\label{pout}
\end{align}
It is easy to see that the possible $P'_{out}$ will be the tensor products that have $I$ or $\sigma_z$ for the qubits that
have  $+1$ in $\bar{s}_{out}$, and $\sigma_x$ or $\sigma_y$ for the other qubits.
There are $D=2^n$ of these products, and then the actual $P_{out}$'s could be obtained by inverting Eq. (\ref{pout}).
Therefore, we are indeed left with an experiment equivalent to the one in Fig. \ref{fig:twirling3} but with $D$ possible intermediary Pauli operators.

The key here is that for two different sets $\mathcal{B}_{J_1}$ and $\mathcal{B}_{J_2}$
corresponding to two different MUBs, there can be only one
$P_{out}$ in common for both. This is because the $D+1$ subsets $\mathcal{B}_J$ are obtained by partitioning the
$D^2-1$ nonidentity Pauli operators into
$D+1$ different subsets of $D-1$ commuting operators. The $\mathcal{B}_J$ are then the generators of these subsets (plus the identity).
Given their properties, any two $\mathcal{B}_{J_1}$ and $\mathcal{B}_{J_2}$, plus commutation relations with them
[Eq. (\ref{thecommutator})],
define a unique operator $P_{out}$ \cite{bendersky_paz_2}.
Moreover, given the nature of the operators involved (Pauli gates and the operations involved in the
change of basis for MUBs), and the number of equations ($n$), $P_{out}$ can be established efficiently \cite{bendersky_paz_2}.

Therefore, if we consider together two experiments $(J_1,m_1,\bar{s}_{out}^{(1)})$ and $(J_2,m_2,\bar{s}_{out}^{(2)})$ [each like
in Fig. \ref{fig:inPairs}(b), with $J_1 \neq J_2$],
there will be only one possible intermediary Pauli gate compatible with both experiments, and it can be
computed in a scalable way.
In practice, we will perform $M$ experiments, and analyzing all the possible $M(M-1)/2$ pairs we will establish
the intermediary Pauli gates that have occurred at least twice, which will be at most $O(M^2)$. Then, we just count the
number of experiments $M_+$ where those operators have potentially occurred among the $D$ possible choices. The corresponding
$\chi_{l,l}$ can be estimated as $(D \chi_{l,l} + 1)/(D+1)=M_+/M$. Notice that $\chi_{l,l}$ is then estimated with an standard
deviation $\leq 1/\sqrt{M}$.
We also recall that $\sum_{l} \chi_{l,l} = 1$, so we can use this to estimate altogether the magnitude of the smaller coefficients.\\

This strategy can be also applied using Clifford gates acting on the initial state instead of using MUB states,
as depicted in Fig. \ref{fig:inPairs}(c).
Again, we use the stabilizer formalism as described before.
Since the Pauli group is the normalizer of the Clifford group, indeed
$ \mathcal{C} P_{k} \mathcal{C}^\dag \cong P_{k'} $ (where $\cong$ means equal up to a global phase).
So this means that the action of $\mathcal{C}_{j}$ on a state is equivalent to changing $(\mathcal{B}_Z, \bar{s})$
to $(\mathcal{B}_P, \bar{s})$, where now $\mathcal{B}_P$ is another
subset of $n$ commuting Pauli operators.
Again we use that the state $(\mathcal{B}_P, \bar{s}_{out})$ can be thought as the result of a Pauli operator $P_{out}$ acting on $(\mathcal{B}_P, \bar{s}_{0})$, which now leaves us in the scheme depicted in Fig. \ref{fig:inPairs}(d). Again, there are $D$ possible operators that
fulfill the requisite of commuting (anticommuting) with the Pauli operators in $\mathcal{B}_P$ that have a corresponding $+1$ ($-1$) in $\bar{s}_{out}$. The argument is completely analogous to the one for the MUB twirl.

We thus resort again on combining two experiments. However, the case of two Clifford twirl experiments is not as simple as the MUB twirl one.
It no longer holds that given two experiments there is one single possible intermediary Pauli gate, because two different Clifford gates may map
$\mathcal{B}_Z$ to two subsets $\mathcal{B}_{P_1}$ and $\mathcal{B}_{P_2}$ that generate two Pauli subgroups that have some operators in common. So not every pair of experiments,
even if $\mathcal{C}_1 \neq \mathcal{C}_2$, will be useful toward establishing the $\chi_{l,l}$ above the threshold of $2/M$.
In practice, we should determine the $2 \times n$ operators $\mathcal{C}_{k} \sigma_z^{(j)} \mathcal{C}^\dagger_{k}$ (where $k=1,2$ are two
randomly chosen Clifford gates) and check whether they constitute two independent sets of generators. If that is the case,
then there is indeed a unique intermediary Pauli gate, as it is always the case with the MUB twirl.
And thanks to the Gottesman-Knill theorem, this can be done efficiently with a classical computer.

To compare both methods, we consider the probability of successfully determining a unique intermediary Pauli gate
given two different experiments drawn from a pool of $M$ experiments.
In the case of the MUB twirl, the probability of success is $\mathcal{P}_{MUB} = D/(D+1)$, since there are $D+1$ possible
values of $J_1$, and having randomly withdrawn one, there are $D$ different ones we could withdraw for $J_2$.

\begin{figure}[h]
\begin{center}
\includegraphics[width=0.37\textwidth, angle=0]{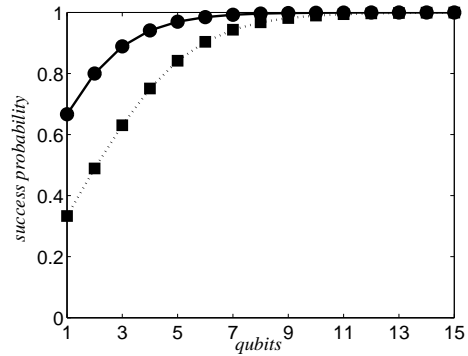}
\end{center}
\caption{Success probability of the two methods: {\large $\bullet$} $\mathcal{P}_{MUB}$, using an MUB twirl;
{\scriptsize $\blacksquare$} $\mathcal{P}_{C}$, using a Clifford twirl.}
\label{fig:prob}
\end{figure}

For the Clifford twirl case, this probability can be calculated as the probability that, given two randomly chosen maximal Abelian Pauli subgroups, the only common element in both groups is the identity, up to a phase. To compute this probability we proceed as follows. We fix the first maximal Abelian group and then compute the probability of adding one by one the operators belonging to the second group. The fixed group has, up to a phase, $D-1$ non identity Pauli operators. If we randomly choose a non identity Pauli operator, namely $P_1$, what is the probability of it not belonging to the fixed group? It is straightforward to see that this probability is $\frac{D^2-D}{D^2-1}$. Now, from the Pauli operators that commute with $P_1$,
what is the probability of picking one Pauli operator $P_2$ that does not belong to the first group? Again, there are a total of $D^2/2-2$ Pauli operators which commute with $P_1$ and are neither $P_1$ nor the identity, but $D/2-1$ of those belong to the first group. So the probability of this happening is $\frac{D^2/2-1-D/2}{D^2/2-2}$. We proceed in the same way, computing the probability of picking a Pauli operator that does not belong either to the first group nor to the group generated by the previously chosen operators. The product of all those probabilities is the probability $\mathcal{P}_{C}$ of having only one intermediary Pauli operator given two Clifford twirl experiments:
\begin{align}
\mathcal{P}_C = \prod_{j=0}^{n-1}  \frac{D^2/2^j-2^j-D/2^j }{D^2/2^j-2^j}
\end{align}
As shown in Fig. \ref{fig:prob}, this probability is smaller but asymptotically equivalent to $\mathcal{P}_{MUB}$.
For the experiments that are being done nowadays, with only a few qubits, the MUB twirl still requires much fewer experimental runs to obtain the larger coefficients.

At this point we can conclude that the method introduced in \cite{bendersky_paz} is indeed the most practical one, and that it can retrieve
the largest diagonal elements of the $\chi$-matrix in the sense that they are above the threshold of $2/M$, for $M$ realizations of the
MUB twirl. This is done in a scalable way, and with a standard deviation $\leq 1/\sqrt{M}$.

This protocol has been experimentally implemented recently,
with photons, to characterize maps on a one-qubit space \cite{schmiegelow_larotonda}. \\

Nevertheless, although efficient, the method demands the errorless implementation of the $\mathcal{V}_{J,m}$ gates
(or of the equally demanding Clifford gates) -- at least relatively errorless compared with any errors in the implementation of $\Lambda$.
If we have a functional quantum device that implements the Hadamard, phase, and controlled-NOT (CNOT) gates
(which are the gates required to implement the Clifford
gates \cite{dankert_thesis} or the MUB states \cite{bendersky_paz_2}) and the Pauli operators
with enough accuracy, we will be in position to study more complex maps with twirls in U($D$).
If however we are still aiming to study gates and sequences whose complexity
is comparable to the one of a Clifford gate in U($D$), this method is unsuitable.
In this case, a more practical alternative arises from the combination of the methods
presented in \cite{emerson_silva} and \cite{lopez_levi} - which will be the object of the next Section.
This proposal allows us to establish the diagonal elements of the $\chi$-matrix coarse-grained
in direction. Indeed, this information is particularly useful when seeking information
for quantum error correction codes, where the particular type of
error ($\sigma_x$, $\sigma_y$ or $\sigma_z$) is irrelevant.

The following method is experimentally quite less demanding, since it requires a twirl in U($2$)$^{\otimes n}$ rather
than one in U($2^n$). On the other hand, as we will see, it assumes a certain structure in the map under
study.
An example of such a scenario
was explicitly shown in the experimental work in \cite{lopez_levi}, using a liquid-state nuclear magnetic resonance (NMR) processor with four qubits.
A relatively large number of qubits easily shows the significative difference between required resources;
to our knowledge, this is the largest number of qubits on which a complete \cite{schmiegelow_larotonda,complete_qpt} or
partial \cite{emerson_silva} quantum process characterization has been attempted.

\section{Methods using a one-qubit twirling of the map under study}
\label{sec:1qubitTwirl}

In this section we concentrate on methods based on a one-qubit twirling of a map \cite{emerson_silva,lopez_levi,levi_lopez}.
That is, the twirl
is a tensor product of twirling operators $U$ acting on each qubit. The two protocols
presented in \cite{emerson_silva} and \cite{lopez_levi} can be actually merged into one.
We review the compact method by proving it all together, which also shows clearly the simplicity and economy of its
implementation (since both \cite{emerson_silva} and \cite{lopez_levi} include their experimental implementation).

In Sec. \ref{sec:twirl} we highlighted the promising role of the Haar twirl. This for example motivated
the first works \cite{emerson_alicki, levi_lopez}. However, as mentioned before, the work by Dankert et al. \cite{dankert} pointed
out an equivalence between a Haar twirl and a Clifford twirl.

Rather than starting from the Haar twirl and crossing over to the Clifford twirl,
we will work directly with the Clifford gates and prove everything from scratch.
For this we will use that the Clifford operators can in turn be decomposed into Pauli operators (the normalizer of the Clifford group)
and the so-called Symplectic operators (the resulting quotient group).\\

We will follow the notation of \cite{emerson_silva}.
Each index $l$ carries the following information: $w$, $\nu_w$, $\mathbf{i}_w$. $w$ is the Pauli weight of
$P_l$, that is, how many of the factors in $P_l$ are nonidentity. The index $\nu_w$ in $\{ 1, \ldots, {n \choose w}\}$ counts the number
of distinct ways that $w$ nonidentity Pauli operators can be distributed over the $n$ factor spaces. The index $\mathbf{i}_w$ is a vector of length $w$
of the form $\mathbf{i}_w=(i_1, i_2,\ldots,i_w)$ with each component being $1=x$, $2=y$, or $3=z$ to denote which Pauli matrix occupies that
respective factor position in the tensor product forming $P_l$. There are $3^w$ of these $\mathbf{i}_w$ for given $w$ and $\nu_w$.\\

We start first with a Pauli twirl (PT) of the map. Thus $\Lambda$ becomes
\begin{align}
\mathbf{\Lambda}^{\mathrm{PT}}(\rho) &= \frac{1}{D^2} \sum_{m=0}^{D^2-1} P_m \mathbf{\Lambda}(P_m \rho P_m) P_m \\
&= \frac{1}{D^2} \sum_{m=0}^{D^2-1} \sum_{l,l'}^{D^2-1} \chi_{l,l'} P_m P_l P_m \rho P_m P_{l'} P_m \\
&= \sum_{l=0}^{D^2-1} \chi_{l,l} P_l \rho P_{l}
\end{align}
This result was proven in \cite{dankert_thesis}. It can be also seen as follows: For $l=l'$, $P_m P_l P_m \rho P_m P_{l} P_m = P_l \rho P_{l}$
since each $P_l$ either commutes or anticommutes with each $P_m$. And if $l=l'$, for each $j$-th factor in which they differ,
we have $P^{(j)}_m P^{(j)}_l P^{(j)}_m \rho P^{(j)}_m P^{(j)}_{l'} P^{(j)}_m = \pm P^{(j)}_l \rho P^{(j)}_{l'}$, with each sign happening for half of the four possible $P^{(j)}_m$. Thus they cancel out in the sum.

We consider now a Symplectic one-qubit twirl (S1T) of the form
\begin{eqnarray}
\mathbf{\Lambda}^{\mathrm{S1T}}(\rho) &=& \frac{1}{3^n} \sum_{m=1}^{3^n} S^\dag_m \mathbf{\Lambda}(S_m \rho S^\dag_m) S_m \\
S_m &=& \bigotimes_{j=1}^n S_m^{(j)} \label{expansionS}
\end{eqnarray}
where each $S_m^{(j)}$ is an element of the set given by $\{\exp(-i(\pi/4)\sigma_p),\  p=x,y,z\}$.
It is straightforward to show that
\begin{eqnarray*}
\frac{1}{3} \sum_{m=1}^3 S^{(j)\dag}_m \sigma_j S^{(j)}_m \rho S^{(j)\dag}_m \sigma_j S^{(j)}_m
= \frac{\sigma_x \rho \sigma_x + \sigma_y \rho \sigma_y + \sigma_z \rho \sigma_z}{3}
\end{eqnarray*}
so after a Clifford (Pauli+Symplectic) one-qubit twirl (C1T) \cite{footnote1} we get
\begin{eqnarray}
\mathbf{\Lambda}^{\mathrm{C1T}}(\rho) = \frac{1}{3^n} \sum_{m=0}^{3^n} S^\dag_m \mathbf{\Lambda}^{\mathrm{PT}}(S_m \rho S^\dag_m) S_m \\
= \sum_{w=0}^{n} \sum_{\nu_w}^{{n \choose w}} \frac{\chi^{col}_{w,\nu_w}}{3^w}
\left( \sum_{\mathbf{i}_w} P_{w,\nu_w,\mathbf{i}_w} \rho P_{w,\nu_w,\mathbf{i}_w} \right)
\end{eqnarray}
where the collective coefficients $\chi^{col}_{w,\nu_w}$ are just the diagonal $\chi$-matrix coefficients $\chi_{l,l}$, re-labeled
$\chi_{w,\nu_w,\mathbf{i}_w}$, after disregarding
(averaging over) the information given by $\mathbf{i}_w$:
\begin{eqnarray}
\chi^{col}_{w,\nu_w} \equiv \sum_{\mathbf{i}_w} \chi_{w,\nu_w,\mathbf{i}_w}
\label{chicol}
\end{eqnarray}
This is so far what was presented in \cite{emerson_silva}, which can also be proven as in \cite{levi_lopez,lopez_levi}
using a different set of tools to handle the Clifford twirl as a Haar twirl \cite{dankert,haartools,levi_lopez}.

Consider the computational state basis $|\bar{v}_h\rangle$, where
$\bar{v}_h$ is a boolean vector of length $n$ and Hamming weight $h$.
(The Hamming weight $h$ of a computational state is just the number of ones appearing in its binary representation.)
The first result we can obtain is that the fidelity of a state $|\bar{v}_h\rangle$
undergoing this transformation is
independent of the actual state,
\begin{eqnarray}
& &f(\Lambda^{\mathrm{C1T}},|\bar{v}_h\rangle)= \Tr[|\bar{v}_h\rangle \langle \bar{v}_h| \mathbf{\Lambda}^{\mathrm{C1T}}(|\bar{v}_h\rangle \langle \bar{v}_h|)] \label{r1}\\
&=& \sum_{w=0}^{n} \sum_{\nu_w}^{{n \choose w}} \frac{\chi^{col}_{w,\nu_w}}{3^w}
\left( \sum_{\mathbf{i}_w}^{3^w} | \langle \bar{v}_h | P_{w,\nu_w,\mathbf{i}_w} |\bar{v}_h\rangle |^2 \right) \label{r2}\\
&=& \sum_{w=0}^{n} \sum_{\nu_w}^{{n \choose w}} \frac{\chi^{col}_{w,\nu_w}}{3^w}
\left( \sum_{\mathbf{i}_w}^{3^w} \langle 0 | P_{w,\nu_w,\mathbf{i}_w} |0\rangle |^2 \right) \label{r3} \\
&=& \sum_{w=0}^{n} \sum_{\nu_w}^{{n \choose w}} \frac{\chi^{col}_{w,\nu_w}}{3^w} \label{r4}
\end{eqnarray}
To go from (\ref{r2}) to (\ref{r3}), we only need to realize that any computational state $|\bar{v}_h\rangle$ is a result of applying
a Pauli operator $P^{\bar{v}_h}_X$ (that has $\sigma_x$ where $\bar{v}_h$ has ones and nonidentity factors otherwise)
to $|0\rangle$. This $P^{\bar{v}_h}_X$ will either commute or anticommute
with $P_{w,\nu_w,\mathbf{i}_w}$ (and the $\pm$ will be absorbed by the modulus squared).
The last equality (\ref{r4}) is obtained by realizing that the only nonidentity $P_{w,\nu_w,\mathbf{i}_w}$ that takes $|0\rangle$ back to it (up
to a global phase) is the Pauli operator that has $\sigma_z$ in all the positions indicated by $\nu_w$ (and thus only one of all the possible
$\mathbf{i}_w$ given $\nu_w$ and $w$).

We must notice that although $f(\Lambda^{\mathrm{C1T}},|\bar{v}_h\rangle)$ is then equivalent to the average fidelity $F(\Lambda^{\mathrm{C1T}})$ of the
process $\Lambda^{\mathrm{C1T}}$, this is not the
average fidelity of the process under study, namely $F(\Lambda)= (D \chi_{0,0} + 1)/(D+1)$ (c.f. Sec. \ref{sec:fullTwirl}).
However, this weaker twirl gives a different insight into the map structure.
The first result we point out, presented in \cite{emerson_silva}, is that we can obtain the diagonal elements of the $\chi$-matrix
grouped by Pauli weight
\begin{align}
p_w &\equiv \sum_{\nu_w}^{{n \choose w}} \sum_{\mathbf{i}_w} \chi_{w,\nu_w,\mathbf{i}_w} = \sum_{\nu_w}^{{n \choose w}} \chi^{col}_{w,\nu_w}
\label{pw}
\end{align}
The parameters $p_w$ and $\chi^{col}_{w,\nu_w}$ are just a coarse-graining of the diagonal elements of the $\chi$-matrix.
The $p_w$ relate to the probability $\Prob{\bar{v}_h,h}$ of obtaining any state $|\bar{v}_h\rangle$
with Hamming weight $h$ when measuring the final state $\Lambda^{\mathrm{C1T}}(|0\rangle \langle 0|)$.
We have
\begin{eqnarray*}
\Prob{\bar{v}_h,h}&=& \Tr[|\bar{v}_h\rangle \langle \bar{v}_h| \Lambda^{\mathrm{C1T}}(|0\rangle \langle 0|)] \\
&=& \sum_{w=0}^{n} \sum_{\nu_w}^{{n \choose w}} \frac{\chi^{col}_{w,\nu_w}}{3^w}
\left( \sum_{\mathbf{i}_w}^{3^w} \langle 0 | P_{w,\nu_w,\mathbf{i}_w} |\bar{v}_h\rangle |^2 \right)
\end{eqnarray*}
For $\langle 0 | P_{w,\nu_w,\mathbf{i}_w} |\bar{v}_h\rangle$ to be nonzero (i.e., $\pm 1$),
$\nu_w$ must indicate nonidentity factors at least where there are
ones in $\bar{v}_h$ (so it must be $w \geq h$). Also the $i_j$ in $\mathbf{i}_w$ must be $1=x$ or $2=y$ for the qubits with ones in $\bar{v}_h$, and $3=z$ for the $w-h$ qubits that have zeros in $\bar{v}_h$ but have a nonidentity factor $P_{w,\nu_w,\mathbf{i}_w}$. There will be
exactly $2^h$ of these operators for given $w$ and $\nu_w$, so
\begin{eqnarray}
\Prob{\bar{v}_h, h}
&=& \sum_{w=h}^{n} \sum_{\nu_w^*=1}^{{n-h \choose w-h}} \frac{2^h}{3^w} \chi^{col}_{w,\nu_h+\nu^*_w}
\label{eq_pwAndWhich}
\end{eqnarray}
where $\nu_h$ indicates a $\chi^{col}_{w,\nu_w}$ for Pauli operators that have a nonidentity factor for at least all the qubits whose corresponding
component in $\bar{v}_h$ is a one. $\nu_w^*$ labels the ${n-h \choose w-h}$ coefficients with $w \geq h$ that fulfill this condition.
If we now discard the ``which qubit'' information given by $\bar{v}_h$, summing over
all the ${n \choose h}$ possibilities, then
\begin{eqnarray}
\Prob{h} &=& \sum_{\bar{v}_h} \Prob{\bar{v}_h,h} \\
&=& \sum_{w=h}^{n} \frac{2^h}{3^w} \sum_{\nu_w^*=1}^{{n-h \choose w-h}} \sum_{\nu_h=1}^{{n \choose h}} \chi^{col}_{w,\nu_h+\nu_w^*} \\
&=& \sum_{w=h}^{n} \frac{2^h}{3^w} {w \choose h} \left( \sum_{\nu_w=1}^{{n \choose w}} \chi^{col}_{w,\nu_w} \right) \\
&=& \sum_{w=h}^{n} \frac{2^h}{3^w} {w \choose h} p_w \label{eq_pw}
\end{eqnarray}
In this way, all the $p_w$ are related to the probabilities of measuring an outcome with Hamming weight $h$
by a $n\times n$ matrix $R_{h,w}=\frac{2^h}{3^w} {w \choose h}$, as stated in \cite{emerson_silva}.\\

We can also keep the ``which qubit'' information and use the probabilities $\Prob{\bar{v}_h, h}$ constructively
to gain even more detail. This strategy was already suggested in \cite{lopez_levi} but more oriented to ensemble quantum information
processors. We present it now in a different manner so it can
be combined with the previous strategy.

Let us replace the descriptors $w$ and $\nu_w$ by $\bar{v}_w$, a boolean vector of length $n$ and Hamming weight $w$ characterizing a Pauli operator $P_l$. $\bar{v}_w$ has a zero in the $j$-th position
if and only if $P_l^{(j)}=I$, otherwise it has a one. For example, the operator $\sigma_z^{(1)}\sigma_x^{(3)}$ for $n=4$ qubits has $\bar{v}_2=(1,0,1,0)$.
There are of course $\sum_{w=0}^n {n \choose w}= 2^{n} = D$ of these vectors describing the $P_l$.

If we use Eq. (\ref{eq_pwAndWhich})
and start with the probability of having all the qubits flipped in the outcome, and go backward
toward the survival probability (i.e., none of the qubits flipped), we find
\begin{subequations} \label{probschi}
\begin{align}
\Prob{n} &= \frac{2^n}{3^n} \chi^{col}_{\bar{v}_n}  \label{chie1}\\
\Prob{\bar{v}_{n-1}, n-1}
&= \frac{2^{n-1}}{3^{n-1}} \chi^{col}_{\bar{v}_{n-1}} + \frac{2^{n-1}}{3^n} \chi^{col}_{\bar{v}_n} \phantom{morespace} \label{chie2}\\
\Prob{\bar{v}_{n-2}, n-2}
&= \frac{2^{n-2}}{3^{n-2}} \chi^{col}_{\bar{v}_{n-2}} \nonumber \\
&+ \sum_{\bar{v}_{n-1}} \frac{2^{n-2}}{3^{n-1}} \chi^{col}_{\bar{v}_{n-1}} + \frac{2^{n-2}}{3^n} \chi^{col}_{\bar{v}_n} \label{chie3}\\
\ldots \ etc. \nonumber
\end{align}
\end{subequations}
So essentially we could determine $\chi^{col}_{\bar{v}_n}$ using (\ref{chie1}), then insert it in (\ref{chie2}) and obtain the $n$
possible $\chi^{col}_{\bar{v}_{n-1}}$ from the different $\Prob{\bar{v}_{n-1}, n-1}$, and then insert that in (\ref{chie3}), and so on and so forth.
These equations define a triangular matrix that relates the probabilities $\Prob{\bar{v}_{h}, h}$
to the collective coefficients $\chi^{col}_{\bar{v}_w}$.
Notice there is no need to perform different experiments to obtain the different probabilities: We only need to implement $M$ realizations of the twirl and keep the outcome of the measurement for each of the realizations.
This outcome should be a $n$-bit string indicating whether each $j$-th qubit was found in $|0\rangle_j$  or $|1\rangle_j$. \\

The problem arises not in obtaining the experimental information, but in its posterior processing. The matrix given by eqs.
(\ref{probschi}) is of size $D\times D$, therefore the cost of the processing would scale exponentially in $n$.
For this strategy to work, it is key to relate it hierarchically to the
determination of the $p_w$: The experimental information required is the same and can be obtained efficiently by sampling. The idea goes as follows.
If we are analyzing a map $\Lambda$
that is close to the identity (a noise channel) or a quantum gate involving a few qubits (typically one or two), then we
would expect that above a certain cut-off Pauli weight $w_{co}$, the $p_w$ will be null.
This is a reasonable expectation: Since $\sum_{w=0}^n p_w = 1$ (the trace-preserving condition), the
$p_w$ cannot all be arbitrarily large, and thus it will be
possible to bound the coefficients above the cut-off by a negligible amount.
In this scenario, the matrix relating the $\Prob{\bar{v}_{h}, h}$ with the $\chi^{col}_{\bar{v}_w}$
will have a size $M_{co} \times M_{co}$, $M_{co} = \sum_{m=0}^{w_{co}} {n \choose m}$, which scales
polynomially in $n$ \cite{footnote2}.
There is a second caveat though. As explained in \cite{emerson_silva,lopez_levi} respectively,
the errors in determining the $p_w$ or the $\chi^{col}_{w,\nu_w}$ scale inefficiently with $w$,
a consequence of the matrices relating them with the corresponding probabilities (eqs. (\ref{eq_pw}) and (\ref{probschi}) respectively).
Although the measured probabilities will have a standard deviation $\leq 1/\sqrt{M}$, this error will propagate into the
$p_w$ or the $\chi^{col}_{w,\nu_w}$ with a factor that grows polynomially with $n$ but exponentially with $w$.
Again, we must resort to neglecting the $p_w$ after a certain cut-off.
The system can be arbitrary large (arbitrary $n$), and as long as the $p_w$ are negligible
above a certain $w_{co}$ (with $w_{co}$ independent of or scaling efficiently with $n$) we will
be able to obtain all the non-negligible $\chi_{w,\nu_w}^{col}$ efficiently.\\

Notice that, in the previous section, the requirement that only a few ($<< D$) coefficients $\chi_{w, \nu_w, \mathbf{i}_w}$
are non-negligible is not a priori. We can indeed run the protocol, efficiently, and arrive at this conclusion.
However, the one-qubit twirling method poses a stronger condition,
since the values of the $\chi^{col}_{w, \nu_w}$ must respond to
a hierarchy associated to their Pauli weights. Only then we can establish the Pauli weight cut-off and run the protocol
[in particular, solve the system of equations (\ref{probschi})].

With the twirl in U($D$) we obtain the coefficients directly with a standard deviation $\leq 1/\sqrt{M}$,
while with the twirl in U($2$)$^{\otimes n}$ we only obtain probabilities $\Prob{\bar{v}_h,h}$ with standard deviations $ \leq 1/\sqrt{M}$, 
which still need to be propagated in order to obtain the estimated error for the $\chi^{col}_{w,\nu_w}$.

With the protocol of Sec. \ref{sec:fullTwirl} \cite{bendersky_paz, bendersky_paz_2},
the measurement of the largest $\chi_{l,l}$ can be done then more precisely, with no coarse-graining
and with no restrictions on the map under study.
Clearly, the protocol of this section \cite{emerson_silva,lopez_levi} is quite less demanding, requiring the implementation
of only $12 n$ one-qubit gates instead of $O(n^2)$ one-qubit and CNOT gates. However, this advantage
is counterbalanced: We have a more restricted and less precise tomographic method.
In practice, nevertheless, the choice between the two will be given by the extent to which we can control our system experimentally.

Finally, we must notice that in both approaches, the methods are universal in the sense that they do not require any prior knowledge on
the specific dynamics of $\Lambda$. The protocol twirling in full space is valid for any linear Hermitian map, while the one with one-qubit twirling
only has the extra requirement of having a structure with a cut-off Pauli weight.
For an example of a characterization incorporating substantial prior knowledge of the dynamics or specific models for $\Lambda$, see \cite{walmsley_kosut}.

\section{The relevance of the diagonal of the $\chi$-matrix}
\label{sec:diagonal}

If we diagonalize the $\chi$-matrix, we will obtain the weights of an operator-sum representation, where the operators
in the sum are the corresponding basis
where the $\chi$-matrix is diagonal. Of course, this basis will not necessarily be the Pauli
operator basis, but in principle a combination of them.
Using the notation of Sec. \ref{sec:chi}, take $\chi=R^\dag S R$ to be the diagonalization
of the $\chi$-matrix written in the Pauli operator basis.
Let $R$ be the change of basis, so
\begin{align*}
\mathbf{\Lambda}(\rho) &= \sum_{m=0}^{D^2-1}  S_{m,m} A_m \rho A^\dag_m \ \ \ \ \ \ \
A_m = \sum_{l=0}^{D^2-1} R^*_{m,l} P_{l}
\end{align*}
where the $A_m$ form an orthonormal basis,
but just as in an operator-sum representation, they are not necessarily unitary (otherwise any process would be unital)
nor Hermitian.
And, as we already mentioned, the $S_{m,m}$ are real but could be negative in principle.
Thus in general neither the $\chi_{l,l}$ nor even the $S_{m,m}$ have a simple interpretation.\\

Nevertheless, despite the different ways of describing the process under study $\Lambda$ in
\cite{dankert,emerson_silva,bendersky_paz,bendersky_paz_2,lopez_levi}, in all the cases they
determine specifically
the diagonal elements of
the $\chi$-matrix of the map in the generalized Pauli operator basis. Notice that either the one-qubit twirl
or the full-space twirl implies a Pauli twirl (since the Pauli operators are a subgroup of the Clifford group
in both cases), and that the Pauli twirl erases the information of the off-diagonal elements of the $\chi$-matrix.
We ask then, what is the meaning of the diagonal? It was assumed in \cite{emerson_silva} that the
$p_w$ represented the probability of an operator of Pauli weight $w$ happening in the process described by $\Lambda$.
In \cite{lopez_levi}, the $\chi^{col}_{w,\nu_w}$ were regarded as indicators of the locality or range of the process,
that is, the probability of an operator involving the qubits in $\nu_w$ happening. These are both quantities that are
relevant to quantum error correction and fault-tolerant quantum computing.\\

Both these interpretations are fair
when the $\chi$-matrix in the Pauli operator basis is approximately diagonal, at least block-diagonal
in blocks characterized by $w$, $\nu_w$.
But that is not generally the case, in particular for maps that will be of our interest -- such as quantum computing gates.
For example, the CNOT gate for qubits $a$ and $b$ has a $\chi$-matrix with only a $4\times4$ nonzero block,
\begin{displaymath}
\chi_{CNOT} = 0.25 \left(
\begin{tabular}{cccc}
1 & 1 & 1 & -1\\
1 & 1 & 1 & -1\\
1 & 1 & 1 & -1\\
-1 & -1 & -1 & 1
\end{tabular}
\right)
\end{displaymath}
corresponding to $P_l=I,\sigma^{(a)}_z,\sigma^{(b)}_x,\sigma^{(a)}_z \otimes\sigma^{(b)}_x$.
Clearly, the off-diagonal coefficients carry critical information with equal weight,
which for example differentiates the CNOT from
 a depolarizing channel with the same $P_l$.

Thus previous
interpretations of $p_w$ and $\chi^{col}_{w,\nu_w}$ are arguable: We could even have in principle
a process involving
a set of qubits given by $w$, $\nu_w$ that has $\chi_{l,l'} \neq 0$ in that block but $\chi_{w,\nu_w}^{col}=0$ in the diagonal.
However, as demonstrated in Secs. \ref{subsec:CP} and \ref{subsec:P},
it is possible to draw a relation between the diagonal and off-diagonal elements of
the $\chi$-matrix.

For CP maps, Eq. (\ref{boundCP}) guarantees that if either $\chi_{l,l}=0$ or $\chi_{l',l'} = 0$, the off-diagonal $\chi_{l,l'}$
is null.
And for positive maps in general, Eq. (\ref{boundP}) gives us a bound that is exponentially close to this result.
This a very powerful result, since once we have established the nonzero diagonal elements, in
order to perform a full characterization we
only need to worry about the off-diagonal elements that correspond to that resulting block.
This hierarchization of the information could potentially allow for
a complete quantum tomography of the process at a scalable cost -- provided that the number of non-null
matrix elements turns out to be $O(poly(n))$.

It is in order here to point out though
that the work in \cite{bendersky_paz,bendersky_paz_2} also presents a strategy to measure the off-diagonal
elements of the $\chi$-matrix. However, an ancillary qubit which is not twirled is required for this task.
The ancilla is assumed to be error-free and outside the system we are looking to characterize.
This does not imply an issue when it comes to scalability, since only one qubit ancilla is
required for arbitrary $D$. Nonetheless, it puts this method in a different category
regarding resources and assumptions when it comes to its implementation.\\

\section{Conclusions}

By revisiting previous work \cite{emerson_silva, bendersky_paz, lopez_levi} we have stated two scalable
approaches for characterizing the diagonal elements of the $\chi$-matrix
in the Pauli operator basis, for any arbitrary quantum process.
We emphasize once more that the work in Secs. \ref{sec:fullTwirl} and \ref{sec:1qubitTwirl} 
arises from the revision of these previous results and goes beyond, which we would like to summarize here:
The general approach discussed in Sec. \ref{sec:fullTwirl} restates the method originally presented in \cite{bendersky_paz}, further clarifying its ability
to measure the largest diagonal elements of the $\chi$-matrix together. We study this protocol by recognizing its familiarity with other twirling 
methods and present a natural alternative approach which, we conclude, is slightly less convenient if we work with only a few qubits.
On the other hand, the approach discussed in Sec. \ref{sec:1qubitTwirl} combines the two protocols originally presented in
\cite{emerson_silva} and \cite{lopez_levi}, by building and again proving both protocols, but simultaneously. 

Furthermore, we have analyzed the two general approaches comparatively, 
establishing their advantages and disadvantages: While one is more powerful, 
the other is more realistic from the implementation point of view.

We have made the point that there are different ways of twirling that reproduce Eqs.
(\ref{haarU}) and (\ref{haarState}). Moreover, we have shown that a
deeper analysis may lead to advantages of one form of twirl over another, in particular for
working with a small number of qubits.
This is the case in Sec. \ref{sec:fullTwirl} when comparing the Clifford twirl and the MUB twirl in U($D$).
Another example of this, but twirling in U($2$)$^{\otimes n}$, can be found in \cite{lopez_levi, lopez_thesis}, where
it is shown that by carefully choosing
the initial state of a twirl experiment, it is possible to reduce the total number of twirl operators from $12^n$ to $6^n$.\\

On the other hand, in the light of Eqs. (\ref{boundCP}) and (\ref{boundP}), our work establishes the relevance of the diagonal
coefficients. We believe that this type of hierarchization of the information is key to achieve complete tomography in a scalable way.
Since the number of parameters is indeed exponentially large, it is necessary to gather them or find relations among them, and then
design protocols that
will retrieve information about a whole group in one parameter.

The coarse-grained coefficients of Sec. \ref{sec:1qubitTwirl} [Eqs. (\ref{chicol}) and (\ref{pw})]
represent one example of grouping. When a sum of nonnegative elements is
null, we can conclude that all the elements in the sum are null.
On the other hand, the bounding of the off-diagonal elements by the diagonals also gives us a form of grouping. When a diagonal element is
null, we can conclude that all the elements corresponding to that row and column are also null.

If many of the parameters turn out to be null indeed in one shot,
eventually leaving only $poly(n)$ non-negligible ones, these strategies become an efficient way to measure all
the coefficients.
Nevertheless, notice that designing methods that retrieve specific partial information is not a trivial task, even when we assume that we can neglect
all the other parameters. We should continue searching for bounds and relations between the characterization parameters of different types of maps.
Also, we should further pursue the design of scalable methods to measure subgroups of information,
while requiring the protocols to rely experimentally on minimum possible resources.

\section{Acknowledgements}

CCL would like to thank the members of the Physics Department at the University of Buenos Aires for their warm hospitality.
This work was supported in part by the National Security Agency NSA under Army Research Office ARO Contract No. W911NF-05-1-0469.


\begin{thebibliography}{99}
\bibitem{basisE} An example of such kind of basis is the one formed
by the generalized Pauli operators. However, untill Sec. \ref{sec:fullTwirl} we refrain from using a particular basis so our results
remain general.
\bibitem{footnote1} The set $\{e^{-i \pi \sigma_q/4}, e^{-i \pi \sigma_q/4}\sigma_p\}$ with $p,q=x,y,z$
generates only half
of the Clifford group for one qubit (up to a global phase). Nevertheless, this set of 12 operators
is enough to implement the Clifford twirl we need.
\end{thebibliography}
\end{document}